\documentclass[twocolumn,prc,showpacs,showkeys]{revtex4}
\usepackage{amsfonts}
\usepackage{amsmath}
\usepackage{amssymb}
\usepackage{graphicx}
\usepackage{rotating}

\providecommand{\U}[1]{\protect\rule{.1in}{.1in}}
\providecommand{\U}[1]{\protect\rule{.1in}{.1in}}
\providecommand{\U}[1]{\protect\rule{.1in}{.1in}}

\begin{document}

\title{Cooperative internal conversion process}
\author{P\'{e}ter K\'{a}lm\'{a}n\footnote{%
retired, e-mail: kalman@phy.bme.hu}}
\author{Tam\'{a}s Keszthelyi\footnote{%
retired, e-mail: khelyi@phy.bme.hu}}
\affiliation{Budapest University of Technology and Economics, Institute of Physics,
Budafoki \'{u}t 8. F., H-1521 Budapest, Hungary\ }

\begin{abstract}
A new phenomenon, called cooperative internal conversion process, in which
the coupling of bound-free electron and neutron transitions due to the
dipole term of their Coulomb interaction permits cooperation of two nuclei
leading to neutron exchange if it is allowed by energy conservation, is
discussed theoretically. General expression of the cross section of the
process is reported in one particle nuclear and spherical shell models as
well in the case of free atoms (e.g. noble gases). A half-life
characteristic of the process is also determined. The case of $Ne$ is
investigated numerically. The process may have significance in fields of
nuclear waste disposal and nuclear energy production.
\end{abstract}

\volumenumber{number}
\issuenumber{number}
\eid{identifier}
\startpage{1}
\endpage{}
\maketitle

\section{Introduction}

The issue of modifying nuclear processes by surroundings is a question of
primary interest of nuclear physics from the very beginnings. In the last
two decades, investigating astrophysical factors of nuclear reactions of low
atomic numbers, in the cross section measurements of the $dd$ reactions in
deuterated metal targets extraordinary observations were made in low energy
accelerator physics \cite{Raiola1}. The phenomenon of increasing cross
sections of the reactions measured in solids compared to the cross sections
obtained in gaseous targets is the so called anomalous screening effect. A
few years ago a systematical survey of the experimental methods applied in
investigating and of the theoretical efforts for the explanation of the
anomalous screening effect was made \cite{Huke} from which one can conclude
that the full theoretical explanation of the effect is still open. On the
other hand, recently the effect of electron screening on the rate of $\alpha 
$ decay was theoretically investigated and observable effect was predicted 
\cite{Karpeshin}.

However, the best known and investigated process in respect of modifying
nuclear processes by surroundings is the internal conversion process, in
which an atomic electron around an excited nucleus takes away nuclear
transition energy of an electromagnetic multipole transition which otherwise
would be prohibited by angular momentum conservation \cite{Hamilton}.

Motivated by the observations of the anomalous screening effect \cite%
{Raiola1}\ we have searched for physical processes that may effect nuclear
reactions in solid state environment. We theoretically found \cite{kk2} that
the leading channel of the $p+d\rightarrow $ $^{3}He$ reaction in solid
environment is the so called solid state internal conversion process, an
adapted version of ordinary internal conversion process. It was shown \cite%
{kk2} that if the reaction $p+d\rightarrow $ $^{3}He$ takes place in solid
material the nuclear energy is taken away by an electron of the environment
instead of the emission of a $\gamma $ photon.

In the usual internal conversion process only one nucleus is involved.
However, very many pairs of nuclei can be found which may go to state of
lower energy if they could cooperate exchanging e.g. a neutron \cite{Shir}.
Therefore it is worth investigating the way the electronic environment could
lead to such cooperation.

Let us take two initial nuclei $_{Z_{1}}^{A_{1}}X$, $_{Z_{2}}^{A_{2}}Y$ and
two final nuclei $_{Z_{1}}^{A_{1}-1}X$, $_{Z_{2}}^{A_{2}+1}Y$ which may be
formed by neutron exchange from $_{Z_{1}}^{A_{1}}X$, $_{Z_{2}}^{A_{2}}Y$. If
the sum $E_{0i\text{ }}$of the rest energies of the initial nuclei is higher
than the sum $E_{0f\text{ }}$of the rest energies of the final nuclei, i.e.
if $E_{0i\text{ }}-E_{0f\text{ }}=\Delta >0$, then the two nuclei $\left(
_{Z_{1}}^{A_{1}}X\text{ and }_{Z_{2}}^{A_{2}}Y\right) $ are allowed to make
a neutron exchange. $\Delta =\Delta _{-}+\Delta _{+},$ with $\Delta
_{-}=\Delta _{A_{1}}-\Delta _{A_{1}-1}$ and $\Delta _{+}=\Delta
_{A_{2}}-\Delta _{A_{2}+1}$. $\Delta _{A_{1}}$, $\Delta _{A_{1}-1}$, $\Delta
_{A_{2}}$, $\Delta _{A_{2}+1}$ are the energy excesses of neutral atoms of
mass numbers $A_{1}$, $A_{1}-1$, $A_{2}$, $A_{2}+1$, respectively \cite{Shir}%
. As it was stated above there exist very many pairs of nuclei that fulfill
the $\Delta >0$ requirement. So it is a question of how these nuclei can
realize the $_{Z_{1}}^{A_{1}}X$, $_{Z_{2}}^{A_{2}}Y$ $\rightarrow $ $%
_{Z_{1}}^{A_{1}-1}X$, $_{Z_{2}}^{A_{2}+1}Y$ neutron exchange transition.

The process%
\begin{equation}
e+_{Z_{1}}^{A_{1}}X+_{Z_{2}}^{A_{2}}Y\rightarrow e^{\prime
}+_{Z_{1}}^{A_{1}-1}X+_{Z_{2}}^{A_{2}+1}Y+\Delta ,  \label{exchange2}
\end{equation}%
that we are going to call cooperative internal conversion process (CICP),
will be discussed with bound-free electron transition and for atomic state
only. Numerically the case of $Ne$ will be investigated. (In Eq.$\left( \ref%
{exchange2}\right) \ e$ and $e^{\prime }$ denote bound and free electrons,
respectively.)

\section{Transition probability per unit time of CICP in single electron -
single nucleon model}

The CICP can be understood with the aid of standard time independent second
order perturbation calculation of quantum mechanics. There are two
perturbations present which cause the effect and which have to be taken into
account. The first is the electric dipole term $V_{Cb}^{dip}$\ of the
Coulomb interaction between a bound electron and a neutron of an atom of
nucleus $_{Z_{1}}^{A_{1}}X$ in which the effective charge of the neutron $%
q_{n}=-Z_{1}e/A_{1}$ \cite{Greiner}. 
\begin{equation}
V_{Cb}^{dip}=\frac{Z_{1}e^{2}}{A_{1}}\frac{4\pi }{3}x_{1}x_{e}^{-2}%
\sum_{m=-1}^{m=1}Y_{1m}^{\ast }(\Omega _{e})Y_{1m}(\Omega _{1}),  \label{VCb}
\end{equation}%
where $Z_{1}$ and $A_{1}$ are charge and mass numbers of the first nucleus, $%
e$ is the elementary charge, $x_{1}$, $x_{e}$ and $\Omega _{1}$, $\Omega _{e}
$ are magnitudes and solid angles of vectors $\mathbf{x}_{1}$, $\mathbf{x}%
_{e}$ which are the relative coordinates of the neutron and the electron in
the first atom, respectively and $Y_{1m}$ denotes spherical harmonics. The
second is the nuclear strong interaction potential $V_{st}$ of nucleus $%
_{Z_{2}}^{A_{2}}Y$. For this nuclear potential a rectangular potential well
is assumed, i.e. $V_{st}=-V_{0}$ $(x_{2}\leq R_{A_{2}})$ and $V_{st}=0$ $%
(x_{2}>R_{A_{2}})$ where $x_{2}$ is the magnitude of vector $\mathbf{x}_{2}$%
, which is the relative coordinate of the neutron in the second nucleus and $%
R_{A_{2}}$ is its radius.

The initial state, which is composed from the state of a bound electron of
the atom having nucleus $_{Z_{1}}^{A_{1}}X$, the state of a bound neutron of
the atomic nucleus $_{Z_{1}}^{A_{1}}X$ and the states of centers of mass
motion, is changed to first order due to the perturbation $V_{Cb}^{dip}$
according to stationary perturbation calculation as%
\begin{eqnarray}
\psi _{i} &=&[\psi _{i}^{(0)}+\sum_{k}a_{bb,k}\psi _{k,bb}^{(0)}
\label{pszii1} \\
&&+\sum_{k}\int a_{bf,k}\psi _{k,bf}^{(0)}d\nu _{k}+\int a_{ff,k}\psi
_{k,ff}^{(0)}d\nu _{k}]e^{-\frac{iE_{i}t}{\hbar }}  \notag
\end{eqnarray}%
Here $E_{i}$ is the total initial energy, which includes the sum $E_{0i\text{
}}$of the rest energies of the initial nuclei $_{Z_{1}}^{A_{1}}X$, $%
_{Z_{2}}^{A_{2}}Y$. $\psi _{i}^{(0)}$ is the product of the unperturbed
bound electron and neutron states and two plane waves $\psi _{CM,A_{1}}$ and 
$\psi _{CM,A_{2}}$ which describe the motions of the centers of mass of the
atoms having $_{Z_{1}}^{A_{1}}X$ and $_{Z_{2}}^{A_{2}}Y$ nucleus,
respectively. $\psi _{k,bb}^{(0)}$ is the product of other bound electron
and neutron states, $\psi _{CM,A_{1}}$ and $\psi _{CM,A_{2}}$. In $\psi
_{k,bf}^{(0)}$ one of the electron and neutron states is bound and the other
is free which are multiplied by $\psi _{CM,A_{1}}$ (in case of free
electron) or by $\psi _{CM,A_{1}-1}$ (in case of free neutron) and $\psi
_{CM,A_{2}}$, where $\psi _{CM,A_{1}-1}$ describes the motion of the center
of mass of the atom having $_{Z_{1}}^{A_{1}-1}X$ nucleus. The last term is
interesting from the point of view of our process and in $\psi _{k,ff}^{(0)}$
both the electron and the neutron are free and their product is multiplied
by $\psi _{CM,A_{1}-1}$ and $\psi _{CM,A_{2}}$. Accordingly 
\begin{equation}
\delta \psi _{i,free}=\int \frac{V_{Cb,ki}^{dip}}{E_{i}-E_{k}}\psi
_{k,ff}^{(0)}d\nu _{k}e^{-\frac{iE_{i}t}{\hbar }}.  \label{Pszii}
\end{equation}%
The sum of the energy of the free electron, neutron and center of mass
states is $E_{k}$, which contains the sum $E_{0k\text{ }}$of the rest
energies and the state density is $d\nu _{k}$ \cite{Landau}.

Taking into account the interaction $V_{st}$ between a free neutron and an
other nucleus $_{Z_{2}}^{A_{2}}Y$ the second order transition probability
per unit time reads as 
\begin{equation}
W_{fi}=\frac{2\pi }{\hbar }\int \left\vert \int \frac{%
V_{st,fk}V_{Cb,ki}^{dip}}{E_{i}-E_{k}}d\nu _{k}\right\vert ^{2}\delta \left(
E_{i}-E_{f}\right) d\nu _{f}.  \label{Wfi}
\end{equation}%
Here $V_{Cb,ki}^{dip}$ and $V_{st,fk}$ are matrix elements of $V_{Cb}^{dip}$
and $V_{st}$ with states $\psi _{k,ff}^{(0)}$, $\psi _{i}^{(0)}$ and $\psi
_{f}^{(0)}$, $\psi _{k,ff}^{(0)}$, respectively. $\psi _{f}^{(0)}$ is the
product of the free electron state, the bound neutron and free center of
mass state of nuclei $_{Z_{1}}^{A_{1}-1}X$\ and $_{Z_{2}}^{A_{2}+1}Y$. The
quantity $d\nu _{f}$ is the density of the final states of sum energy $E_{f}$%
, which comprises the sum $E_{0f\text{ }}$of the rest energies of the final
nuclei $_{Z_{1}}^{A_{1}-1}X$\ and $_{Z_{2}}^{A_{2}+1}Y$. If the free states
are plane waves of wave vectors $\mathbf{k}_{e}$, $\mathbf{k}_{1}$ and $%
\mathbf{k}_{2}$ corresponding to the wave vectors of the free electron, the $%
_{Z_{1}}^{A_{1}-1}X$\ nucleus which has lost the neutron and the nucleus $%
_{Z_{2}}^{A_{2}+1}Y$ which has taken up the neutron then 
\begin{equation}
d\nu _{f}=\frac{V^{3}}{\left( 2\pi \right) ^{9}}d\mathbf{k}_{e}d\mathbf{k}%
_{1}d\mathbf{k}_{2},\text{ }d\nu _{k}=\frac{V}{\left( 2\pi \right) ^{3}}d%
\mathbf{k}_{n},  \label{dnuf}
\end{equation}%
where $\mathbf{k}_{n}$ is the wave vector of the free (intermediate)
neutron, and $V$ is the volume of normalization. The initial wave vectors of
atoms having nuclei $_{Z_{1}}^{A_{1}}X$\ and $_{Z_{2}}^{A_{2}}Y$ are
negligible. (One can say that the initial bound neutron of nucleus $%
_{Z_{1}}^{A_{1}}X$\ is excited into an intermediate free state due to the
dipole term $V_{Cb}^{dip}$ of its Coulomb-interaction with one of the bound
atomic electrons and from this intermediate state it is captured by an other
nucleus $_{Z_{2}}^{A_{2}}Y$ due to its nuclear potential $V_{st}$ created by
strong interaction forming the nucleus $_{Z_{2}}^{A_{2}+1}Y$ in this way.)
The nuclear energy difference $\Delta $, that is the reaction energy, is
distributed between the kinetic energies of the final free electron and the
two final nuclei. All told, in Eq.$\left( \ref{exchange2}\right) $ the
nucleus $_{Z_{1}}^{A_{1}}X$ looses a neutron which is taken up by the
nucleus $_{Z_{2}}^{A_{2}}Y$ forming $_{Z_{2}}^{A_{2}+1}Y$ in this manner.

\section{Cross section of CICP in single electron - single nucleon model}

The cross section $\sigma _{bf}$ of the bound-free $\left( bf\right) $
electron\ transitions of CICP can be determined from the transition
probability per unit time by standard method. The evaluation of the
transition probability per unit time is carried out in one particle nuclear
model. Hydrogen like state of binding energy $E_{Bi}$ and Coulomb-factor
corrected plane wave are used as initial, bound and final, free electron
states. The motion of the intermediate neutron and the two final nuclei are
also described by plane waves. The rest masses of the two initial nuclei of
mass numbers $A_{1}$ and $A_{2}\ $are $m_{1}=A_{1}m_{0}$ and $%
m_{2}=A_{2}m_{0}$ where $m_{0}c^{2}=931.494$ $MeV$ is the atomic energy unit.

When calculating $\sigma _{bf}$, it is reasonable to use the $e^{i\frac{m_{e}%
}{\left( A_{1}-1\right) m_{0}}\mathbf{k}_{1}\cdot \mathbf{x}_{e}}=1$ long
wavelength approximation in the electron part of the Coulomb matrix element
since $\frac{m_{e}}{\left( A_{1}-1\right) m_{0}}\ll 1$. The analysis of $%
\sigma _{bf}$ shows that those processes give essential contribution to the
cross section in which $k_{e}\ll $ $k_{1}$ and $k_{e}\ll $ $k_{2}$ where $%
k_{e}$, $k_{1}$ and $k_{2}$ are the magnitudes of the wave vectors of $%
\mathbf{k}_{e}$, $\mathbf{k}_{1}$ and $\mathbf{k}_{2}$. In this case as a
consequence of momentum conservation the integration over $\mathbf{k}_{1}$
can be carried out resulting the substitution $\mathbf{k}_{1}=-\mathbf{k}_{2}
$. Thus $E_{f}$, which is the sum of the kinetic energies of the free
electron, particle $_{Z_{1}}^{A_{1}-1}X$ and particle $_{Z_{2}}^{A_{2}+1}Y$,
(in the Dirac-delta) and $E\left( k_{e},k_{2}\right) $, which is the sum of
the kinetic energies in the intermediate state (in the energy denominator)
become $E_{f\text{ }}=\hbar ^{2}k_{2}^{2}/\left[ 2m_{0}a_{12}\right] +\hbar
^{2}k_{e}^{2}/\left( 2m_{e}\right) $ with $a_{12}=\left( A_{1}-1\right)
\left( A_{2}+1\right) /\left( A_{1}+A_{2}\right) $ and $E\left(
k_{e},k_{2}\right) =\frac{A_{1}}{A_{1}-1}\hbar ^{2}k_{2}^{2}/\left(
2m_{0}\right) +\hbar ^{2}k_{e}^{2}/\left( 2m_{e}\right) $, respectively (the
intermediate neutron has wave vector $-\mathbf{k}_{2}$). These
simplifications result 
\begin{eqnarray}
\sigma _{bf} &=&\left( \frac{2Z_{1}}{3A_{1}}V_{0}\right) ^{2}\frac{\alpha
_{f}^{2}\hbar c^{2}}{v\left( 2\pi \right) ^{3}}\int \int \int \delta (E_{f%
\text{ }}-\Delta _{Bi})  \label{sigma0bf} \\
&&\times \frac{\left\vert \sum_{m=-1}^{m=1}I_{ef}^{m}\left( \mathbf{k}%
_{e}\right) I_{1}^{m}\left( \mathbf{k}_{2}\right) \right\vert ^{2}\left\vert
I_{2}\left( \mathbf{k}_{2}\right) \right\vert ^{2}d\mathbf{k}_{e}d\mathbf{k}%
_{2}}{\left[ E\left( k_{e},k_{2}\right) +\Delta _{n}-\Delta _{-}+E_{Bi}%
\right] ^{2}},  \notag
\end{eqnarray}%
where $v$ is the relative velocity of the two atoms, $c$ is the velocity of
light (in vacuum),%
\begin{equation}
I_{ef}^{m}\left( \mathbf{k}_{e}\right) =\int u_{i}\left( \mathbf{x}%
_{e}\right) Y_{1m}^{\ast }(\Omega _{e})x_{e}^{-2}e^{-i\mathbf{k}_{e}\cdot 
\mathbf{x}_{e}}d\mathbf{x}_{e},  \label{ief}
\end{equation}%
\begin{equation}
I_{1}^{m}\left( \mathbf{k}_{2}\right) =\int \phi _{i}\left( \mathbf{x}%
_{1}\right) x_{1}Y_{1m}(\Omega _{n1})e^{-i\mathbf{k}_{2}\frac{A_{1}}{\left(
A_{1}-1\right) }\cdot \mathbf{x}_{1}}d\mathbf{x}_{1},  \label{i1m}
\end{equation}%
and%
\begin{equation}
I_{2}\left( \mathbf{k}_{2}\right) =\int_{\Delta V}\phi _{f}^{\ast }\left( 
\mathbf{x}_{2}\right) e^{i\mathbf{k}_{2}\cdot \mathbf{x}_{2}}d\mathbf{x}_{2}.
\label{i2}
\end{equation}%
Here $u_{i}\left( \mathbf{x}_{e}\right) $ and $\phi _{i}\left( \mathbf{x}%
_{1}\right) $ are the initial bound electron and neutron states and $\phi
_{f}\left( \mathbf{x}_{2}\right) $ is the final bound neutron state. $\alpha
_{f}$ denotes the fine structure constant and $\hbar $ is the reduced
Planck-constant. $\Delta _{n}=8.071$ $MeV$ is the energy excess of the
neutron. Furthermore, $m_{e}$ is the rest mass of the electron and $\Delta
_{Bi}=\Delta -E_{Bi}$. $\Delta V$ in $I_{2}\left( \mathbf{k}_{2}\right) $ is
that volume of the surface of the second nucleus (of $A_{2}$) in which
direct neutron capture may be assumed \cite{Blatt}. It can be considered as
a shell of a sphere of radius $R_{A_{2}}$ and of thickness $L$, where $L$ is
the mean free path of the ingoing neutron in the nucleus.

\section{Cross section of CICP in spherical nuclear shell model}

The bound, initial electron, the initial and final nuclear states, which are
used, have the form: $u_{i}\left( \mathbf{x}_{e}\right) =R_{i}\left(
x_{e}\right) Y_{js}(\Omega _{e})$, $\phi _{i}\left( \mathbf{x}_{1}\right)
=\varphi _{i}\left( x_{1}\right) Y_{l_{i}m_{i}}(\Omega _{1})/x_{1}$ and $%
\phi _{f}\left( \mathbf{x}_{2}\right) =\varphi _{f}\left( x_{2}\right)
Y_{l_{f}m_{f}}(\Omega _{2})/x_{2}$. Here $j$\ and $s$ are orbital angular
momentum and magnetic quantum numbers of the bound electron state. $\varphi
_{i}\left( x_{1}\right) /x_{1}$ and $\varphi _{f}\left( x_{2}\right) /x_{2}$
are the radial parts of the one particle shell-model solutions of quantum
numbers $l_{i}$, $m_{i}$ and $l_{f}$, $m_{f}$. In the cases to be
investigated the corresponding part $R_{0\Lambda }=b_{k}^{-1/2}\Gamma
(\Lambda +3/2)^{-1/2}2^{1/2}\rho _{k}^{\Lambda +1}\exp (-\rho _{k}^{2}/2)$
of $0\Lambda $ one particle spherical shell model states \cite{Pal} is
applied as $\varphi _{i}\left( x_{1}\right) $ and $\varphi _{f}\left(
x_{2}\right) $. Here $\rho _{k}=x_{k}/b_{k}$, $b_{k}=\left( \frac{\hbar }{%
m_{0}\omega _{sh,k}}\right) ^{1/2}$ and $\hbar \omega _{sh,k}=41A_{k}^{-1/3}$
(in $MeV$ units, \cite{Greiner}) with $k=1,2$ corresponding to $A_{1}$ and $%
A_{2}$, and $\Gamma (x)$ is the gamma function.

Thus in the case of spherical shell model states 
\begin{eqnarray}
\sigma _{bf,sh} &=&\frac{32}{3v}\left( \frac{Z_{1}}{A_{1}}V_{0}\right)
^{2}\left( 2l_{f}+1\right) \int \left\vert J_{2}^{l_{f}}(k_{2})\right\vert
^{2}  \label{sigma02} \\
&&\times \int \sum_{l,\lambda }\frac{N_{l\lambda }\left\vert J_{1}^{\lambda
}(k_{2})\right\vert ^{2}\left\vert J_{e}^{l}(k_{e})\right\vert ^{2}}{\left[
E\left( k_{e},k_{2}\right) +\Delta _{n}-\Delta _{-}+E_{Bi}\right] ^{2}} 
\notag \\
&&\times \alpha _{f}^{2}\hbar c^{2}\delta (E_{f2\text{ }}-\Delta
_{Bi})k_{e}^{2}dk_{e}k_{2}^{2}dk_{2},  \notag
\end{eqnarray}%
where in case of $0l_{f}$ final nuclear state with $\rho
_{f}=R_{A_{2}}/b_{2} $ 
\begin{equation}
\left\vert J_{2}^{l_{f}}(k_{2})\right\vert ^{2}=\frac{\pi L^{2}\rho
_{f}^{2l_{f}+3}e^{-\rho _{f}^{2}}J_{l_{f}+\frac{1}{2}}^{2}(k_{2}R_{A_{2}})}{%
\Gamma \left( l_{f}+\frac{3}{2}\right) k_{2}}.  \label{Jlf}
\end{equation}%
In the case of $0l_{i}$ initial nuclear state 
\begin{equation}
J_{1}^{\lambda }(k_{2})=\int R_{0l_{i}}\left( x_{1}\right) j_{\lambda }(%
\frac{A_{1}}{A_{1}-1}k_{2}x_{1})x_{1}^{2}dx_{1},  \label{Jnlam}
\end{equation}%
\begin{equation}
J_{e}^{l}(k_{e})=F_{Cb}^{1/2}\left( k_{e}\right) \int R_{i}\left(
x_{e}\right) j_{l}(k_{e}x_{e})dx_{e}\text{ \ and }  \label{Jel}
\end{equation}%
\begin{equation}
N_{l\lambda }=(2l+1)(2\lambda +1)\left( 
\begin{array}{ccc}
j & l & 1 \\ 
0 & 0 & 0%
\end{array}%
\right) ^{2}\left( 
\begin{array}{ccc}
l_{i} & 1 & \lambda \\ 
0 & 0 & 0%
\end{array}%
\right) ^{2}.  \label{Njllam}
\end{equation}%
The parenthesized expressions are Wigner 3j symbols. (The suffix $sh$ of any
quantity denotes that it is calculated in the one particle spherical shell
model.) In $\left( \ref{Jel}\right) $ $F_{Cb}\left( k_{e}\right) $is the
Coulomb factor.

We restrict ourselves to$\ 1s$ initial electronic state of $R_{i}\left(
x_{e}\right) =2a^{-3/2}\exp (-x_{e}/a)$ with $a=a_{0}/Z_{eff}$, where $a_{0}$
is the Bohr-radius, $Z_{eff}=\sqrt{E_{B}/Ry}$ and $Ry$ is the Rydberg energy
and use the $F_{Cb}\left( k_{e}\right) =2\pi /\left( k_{e}a\right) $
approximation. Since $j=0$ and $(2l+1)\left( 
\begin{array}{ccc}
0 & 1 & l \\ 
0 & 0 & 0%
\end{array}%
\right) ^{2}=\delta _{1,l}$, 
\begin{equation}
N_{1\lambda }=(2\lambda +1)\left( 
\begin{array}{ccc}
l_{i} & 1 & \lambda \\ 
0 & 0 & 0%
\end{array}%
\right) ^{2}.  \label{N01la}
\end{equation}%
Keeping the leading term of $J_{e}^{1}(k_{e})$, introducing $k_{2}=k_{0}x$,
and carrying out integration over $k_{e}$ with the aid of the
energy-Dirac-delta and in the case of $l_{i}=even$ [$l_{i}=2$; $%
Ne(3/2^{+},0d)$] to be investigated one obtains 
\begin{eqnarray}
\sigma _{bf,sh} &=&\frac{2^{10}\pi ^{3}}{3v}\frac{Z_{1}^{2}V_{0}^{2}}{%
A_{1}^{2}\hbar ^{2}c}\frac{b_{1}^{5}L^{2}}{\lambdabar _{e}a_{0}^{2}}\frac{%
m_{0}}{m_{e}}a_{12}\left( 2l_{f}+1\right)  \label{sigma03} \\
&&\times \frac{\rho _{f}^{2l_{f}+3}e^{-\rho _{f}^{2}}}{\Gamma \left( l_{f}+%
\frac{1}{2}\right) }\sum_{\lambda =l_{i}\pm 1}\frac{N_{1\lambda }\left(
k_{0}b_{1}\right) ^{2\lambda }}{\Gamma \left( \lambda +\frac{3}{2}\right) }%
S_{\lambda }.  \notag
\end{eqnarray}%
Here $k_{0}=\sqrt{2m_{0}c^{2}a_{12}\Delta _{B}}/\left( \hbar c\right) $, $%
\lambdabar _{e}=\hbar /(m_{e}c)$, 
\begin{equation}
S_{\lambda }=\int_{0}^{1}f\left( x\right) g_{\lambda }\left( x\right) dx,%
\text{ with }x=k_{2}/k_{0},  \label{slam}
\end{equation}%
\begin{equation}
f\left( x\right) =\frac{\left( 1-x^{2}\right) x^{2\lambda
+1}e^{-(k_{0}b_{1})^{2}x^{2}}J_{l_{f}+\frac{1}{2}}^{2}(xk_{0}R_{A_{2}})}{%
\left[ 1+\frac{\Delta _{Bi}}{E_{B}}\left( 1-x^{2}\right) \right] ^{2}\left[ 
\frac{A_{1}a_{12}}{A_{1}-1}x^{2}+1+\xi \right] ^{2}},  \label{fkszi}
\end{equation}%
and $\xi =\left( \Delta _{n}-\Delta _{-}+E_{Bi}\right) /\Delta _{Bi}$. In Eq.%
$\left( \ref{slam}\right) $ $g_{\lambda }\left( x\right) =1$ if $\lambda
=l_{i}+1$ and 
\begin{equation}
g_{\lambda }\left( x\right) =\left( 2l_{i}+1\right) ^{2}-2\left(
2l_{i}+1\right) (k_{0}b_{1}x)^{2}+(k_{0}b_{1}x)^{4}  \label{glak2}
\end{equation}%
if $\lambda =l_{i}-1$.

The differential cross section $d\sigma _{bf,sh}/dE_{2}$ of the process can
be determined with the aid of 
\begin{equation}
P(x)=\sum_{\lambda =l_{i}\pm 1}\frac{N_{1\lambda }\left( k_{0}b_{1}\right)
^{2\lambda }}{\Gamma \left( \lambda +\frac{3}{2}\right) }\frac{f\left(
x\right) g_{\lambda }\left( x\right) }{x}  \label{Gy}
\end{equation}%
as $d\sigma _{bf,sh}/dE_{2}=K_{bf}\left[ P(x)\right] _{x=\sqrt{z}}/\left(
2E_{20}\right) $ where $z=E_{2}/E_{20}$ with $E_{20}=\left( A_{1}-1\right)
\Delta _{Bi}/\left( A_{1}+A_{2}\right) $, which is the possible maximum of
the kinetic energy $E_{2}$ of particle $_{Z_{2}}^{A_{2}+1}Y$ (particle $5$)
created in the process, $K_{bf}$ stands for the whole factor which
multiplies the sum in $\left( \ref{sigma03}\right) $. In the case of $e+$ $%
_{10}^{21}Ne+$ $_{10}^{21}Ne\rightarrow e^{\prime }+$ $_{10}^{20}Ne+$ $%
_{10}^{22}Ne+\Delta $ reaction happening from the $K$ shell, $l_{i}=l_{f}=2$%
, $\Delta =3.603$ $MeV$, $E_{Bi}=870.1$ $eV$ and $\Delta _{-}=1.310$ $MeV$ ($%
2E_{20}=3.26$ $MeV$ and $K_{bf}/\left( 2E_{20}\right) =4.77\times 10^{-25}$ $%
cm^{3}s^{-1}MeV^{-1}/v)$. $d\sigma _{bf,sh}/dE_{2}$ has accountable values
near below $z=1$, i.e. if $E_{2}\sim E_{20}$.

The differential cross section $d\sigma _{bf,sh}/dE_{e}=K_{bf}\left[ P(x)%
\right] _{x=\sqrt{1-z}}/\left( 2\Delta _{Bi}\right) $ can also be determined
with the aid of $P(x)$ where $z=E_{e}/\Delta _{Bi}$, $E_{e}$ is the kinetic
energy of the electron and $K_{bf}$ is defined and is given above. $d\sigma
_{bf,sh}/dE_{e}$ has accountable values near above $z=0$, i.e. if $E_{e}\sim
0$.

\section{Numerical calculation in case of $Ne$}

The transition probability per unit time $\lambda _{1}$ of CICP of one
nucleus of mass number $A_{1}$ of an atomic gas of number density $n$
created by all those isotopes of mass number $A_{2}$ for which CICP is
allowed, reads in the spherical shell model as $\lambda
_{1}=n\sum_{A_{2}}r_{A_{2}}v\sigma _{bf,sh}$ in the $\sigma _{bf}=\sigma
_{bf,sh}$ approximation used. Here $r_{A_{2}}$ is the relative natural
abundance of isotope of mass number $A_{2}$ (in the case of $Ne$ to be
investigated the $A_{2}=A_{1}$ event is only possible). Furthermore $\tau
_{1/2,1}=\ln 2/\lambda _{1}$, which can be considered as a half-life.

Finally, $r_{A_{1}}n\lambda _{1}$ is the rate per unit volume of the sample
produced by the nuclei of mass number $A_{1}$ which take part as initial
nuclei in CICP, and $r_{tot}=\sum_{A_{1}}r_{A_{1}}n\lambda _{1}$, which is
the total rate per unit volume of the sample.

In the numerical calculation $V_{0}=50$ $MeV$ is used \cite{Greiner}. In the
case of $Ne$ only the $e+$ $_{10}^{21}Ne+$ $_{10}^{21}Ne\rightarrow
e^{\prime }+$ $_{10}^{20}Ne+$ $_{10}^{22}Ne+\Delta $ reaction of $\Delta
=3.603$ $MeV$ is allowed. CICP does not work in $Ar$ since all the possible
channels are energetically forbidden. On the other hand in the case of $Kr\ $%
and $Xe$ nuclei the applicability of the spherical shell model may already
be questionable. For $Ne$ the transition probability per unit time is
estimated as $\lambda _{1}>\lambda _{1}(K)$, which is the transition
probability per unit time of the bound-free CICP from the $K$ shell of $Ne$.
The initial and final states of $^{21}Ne$ and $^{22}Ne$ are supposed to be $%
0d$ spherical shell model states of $l_{i}=l_{f}=2$, $%
r_{A_{1}}=r_{A_{2}}=r_{21}=0.0027$, and $\Delta _{-}=1.310$ $MeV$. The
electron binding energy in the $K$ shell is $E_{Bi}=870.1$ $eV$. $v\sigma
_{bf,sh}(K)=2.48\times 10^{-34}$ $cm^{3}s^{-1}$ is obtained in the case of
bound-free CICP from the $K$ shell of $Ne$. Taking this, the $\lambda
_{1}>1.8\times 10^{-17}$ $s^{-1}$ and $\tau _{1/2,1}<3.8\times 10^{16}$ $s$ (%
$1.2\times 10^{9}$ $y$) and $r_{tot}>1.25$ $cm^{-3}s^{-1}$ for a gas of
normal state $\left( n=2.652\times 10^{19}cm^{-3}\right) $. The estimated
half life is so long that the decay through CICP does not alter natural
abundance of $Ne$ in observable measure. However, the rate is high enough to
be measurable. $_{10}^{20}Ne$ and $_{10}^{22}Ne$ are mostly formed with
energy near below $E_{10}=\frac{A_{1}+1}{A_{1}+A_{2}}\Delta _{Bi}=1.97$ $MeV$
and $E_{10}=\frac{A_{1}-1}{A_{1}+A_{2}}\Delta _{Bi}=1.63$ $MeV$ and with
wave vectors of opposite direction. Therefore it is plausible to observe
their creation in coincidence measurement.

\section{Other results and discussion}

Although the obtained $\lambda _{1}$ of $Ne$ is rather small, the
Weisskopf-estimation of the cross section indicate that it, and consequently 
$\lambda _{1}$,\ may be increased very much with the increase of the atomic
number.

\begin{table}[tbp]
\tabskip=8pt 
\centerline {\vbox{\halign{\strut $#$\hfil&\hfil$#$\hfil&\hfil$#$
\hfil&\hfil$#$\hfil&\hfil$#$\hfil&\hfil$#$\cr
\noalign{\hrule\vskip2pt\hrule\vskip2pt}
isotope&\tau (y)&A-1,A+1&\Delta_{-} (MeV)&\Delta_{+} (MeV) \cr
\noalign{\vskip2pt\hrule\vskip2pt}
^{113m}Cd & 14.1 & 112, 114 & 1.795 & 1.235 \cr
^{121m}Sn & 55 & 120, 122 & 1.907 & 0.749 \cr
^{151}Sm & 90 & 150, 152 & 2.475 & 0.186 \cr
^{79}Se & 6.5\times 10^{4} & 78, 80 & 1.108 & 1.842 \cr
^{93}Zr & 1.53\times 10^{6} & 92, 94 & 1.337 & 0.149 \cr
^{107}Pd & 6.5\times 10^{6} & 106,108 & 1.533 & 1.149 \cr
\noalign{\vskip2pt\hrule\vskip2pt\hrule}}}}
\caption{Data for cooperative internal conversion process of long lived
nuclear fission products. (Data to reaction $\left( \protect\ref{exchange2}%
\right) $.) $A-1$ and $A+1$ are the mass numbers of the two final izotopes, $%
\protect\tau $ is the half-life of the fission product in $y$ units. For the
definition of $\Delta _{-}$ and $\Delta _{+}$ see the text. }
\end{table}
Moreover, the magnitudes of $\sigma _{bf}$ and $\lambda _{1}$ are very
sensitive to the model applied, e.g. if neutron capture of nucleus of $A_{2}$
is not restricted to the direct reaction then the integral in $I_{2}\left( 
\mathbf{k}_{2}\right) $ must be carried out over the whole volume of the
nucleus. In this case $\sigma _{bf,sh}$ and $\lambda _{1}$ are increased by
a factor of about $240$. Furthermore, $\lambda _{1}$ can essentially
increase e.g. with the increase of pressure. Therefore, contrary to the
smallness of the value of~$\lambda _{1}$ obtained in the case of $Ne$, CICP
could play a role in the problem of nuclear waste disposal. In Table I. some
long lived fission products are listed which can take part in CICP. The
positive values of $\Delta _{-}$ and $\Delta _{+}$ indicate that each pair
of the listed isotopes can produce CICP. Consequently, it seems to stand a
practical chance to accelerate the decay of the listed isotopes if they are
collected in appropriately high concentration and density in atomic state,
which is, however, a great technical challenge.

\begin{table}[tbp]
\tabskip=8pt 
\centerline {\vbox{\halign{\strut $#$\hfil &\hfil$#$\hfil&\hfil$#$
\hfil&\hfil$#$\hfil\cr
\noalign{\hrule\vskip2pt\hrule\vskip2pt}
isotope&r_{A}&\Delta_{+}(A) (MeV)&\Delta (MeV) \cr
\noalign{\vskip2pt\hrule\vskip2pt}
\text{ }_{4}^{9}Be &1.0& -1.259 &4.587\cr
\text{ }_{5}^{10}B &0.199& 3.382 &9.228\cr
\text{ }_{5}^{11}B &0.812& -4.701 &1.145\cr
\text{ }_{6}^{12}C &0.989& -3.125 &2.721\cr
\text{ }_{6}^{13}C &0.011& 0.105 &5.951\cr
\text{ }_{7}^{14}N &0.99634& 2.762 &8.608\cr
\text{ }_{7}^{15}N &0.00366& -5.581 &0.265\cr
\text{ }_{8}^{16}O &0.99762& -3.928 &1.918\cr
\text{ }_{8}^{17}O &0.00038& -0.027 &5.819\cr
\noalign{\vskip2pt\hrule\vskip2pt\hrule}}}}
\caption{The values of the quantities $\Delta _{+}(A)$ and $\Delta =\Delta
_{-}(dp)+\Delta _{+}(A)$ of the $\text{ }e+d+\text{ }_{Z}^{A}X\rightarrow
e^{\prime }+p+\text{ }_{Z}^{A+1}X+\Delta $ reaction. $r_{A}$ is the natural
abundance of the isotope. $\Delta _{-}(dp)=\Delta _{d}-\Delta _{p}=5.846$ $%
MeV$, $\Delta _{d}$ and $\Delta _{p}$ are mass excesses of $d$ and $p$.}
\label{Table2}
\end{table}

Finally a special family of CICP reactions the $e+d+\text{ }%
_{Z}^{A}X\rightarrow e^{\prime }+p+\text{ }_{Z}^{A+1}X+\Delta $ reaction
family is worth mentioning. The quantity $\Delta _{-}(dp)=\Delta _{d}-\Delta
_{p}=5.846$ $MeV$, which is characteristic of neutron loss of the deuteron.
Here $\Delta _{d}$ and $\Delta _{p}$ are mass excesses of $d$ and $p$. The
energy of the reaction is $\Delta =\Delta _{-}(dp)+\Delta _{+}$ and some
concrete reactions together with their $\Delta _{+}$ and $\Delta $ values
are listed, without completeness, in Table II.. Chances are that these
reactions may open new perspectives in the field of nuclear energy
production.

On the grounds of our results it can be stated that CICP seems to be able to
significantly modify nuclear processes by surroundings.

The authors are indebted to E. Lakatos for the valuable and stimulating
discussions.

\end{document}